\documentclass[aip,jcp,reprint,amssymb,superscriptaddress]{revtex4-1}

\usepackage{graphicx}
\usepackage{dcolumn}
\usepackage{bm}
\usepackage{wasysym}
\usepackage{courier}

\begin{document}


\title{{\Large Modeling solvation effects in real-space and real-time within Density Functional Approaches}}

\author{Alain Delgado}
\affiliation{Center S3, CNR Institute of Nanoscience, Via Campi 213/A, 41125 Modena, Italy.}
\affiliation{Centro de Aplicaciones Tecnol\'ogicas y Desarrollo Nuclear, Calle 30 \# 502, 11300 La Habana, Cuba.}
\author{Stefano Corni}
\affiliation{Center S3, CNR Institute of Nanoscience, Via Campi 213/A, 41125 Modena, Italy.}
\author{Stefano Pittalis}
\affiliation{Center S3, CNR Institute of Nanoscience, Via Campi 213/A, 41125 Modena, Italy.}
\author{Carlo Andrea Rozzi}
\affiliation{Center S3, CNR Institute of Nanoscience, Via Campi 213/A, 41125 Modena, Italy.}

\begin{abstract}
The Polarizable Continuum Model (PCM) can be used in conjunction with Density Functional Theory (DFT) and its time-dependent
extension (TDDFT) to simulate the electronic and optical properties of molecules and nanoparticles immersed in a dielectric
environment, typically liquid solvents. In this contribution, we develop a methodology to account for solvation effects in
real-space (and real-time) (TD)DFT calculations. The boundary elements method is used to calculate the solvent reaction potential
in terms of the apparent charges that spread over the Van der Waals solute surface. In a real-space representation this potential
may exhibit a Coulomb singularity at grid points that are close to the cavity surface. We propose a simple
approach to regularize such singularity by using a set of spherical Gaussian functions to distribute the apparent charges. We
have implemented the proposed method in the {\sc Octopus} code and present results for the electrostatic contribution to the
solvation free energies and solvatochromic shifts for a representative set of organic molecules in water.
\end{abstract}


\maketitle

\section{Introduction}
\label{sec_intro}
Most of the innovative applications investigated in nanosciences rely on the physicochemical properties of novel materials which
can be modulated in a non-trivial manner due to the interaction with surrounding environments like solvent solutions, mesoscopic
nanostructures and surfaces. The necessity of considering dielectric effects on the opto-electronic and dynamical properties of
molecular systems is exemplified by a vast scientific literature addressing applications in different fields such as
dye-sensitized solar cells, \cite{gratzel_solar_energy_2005} water-soluble functionalized fullerens for
biotechnology, \cite{nakamura_funcfullerenes_2003, popov_emf_chemrev_2013} porphyrin-based complexes used as artificial
photosynthetic reaction center \cite{imahori_porphyrin_2004} or as efficient sensitizers for photodynamic therapies,
\cite{paszko_photodynamics_2011} among others.

\begin{figure}[t]
\begin{center}
\includegraphics[width=0.9 \linewidth,angle=0]{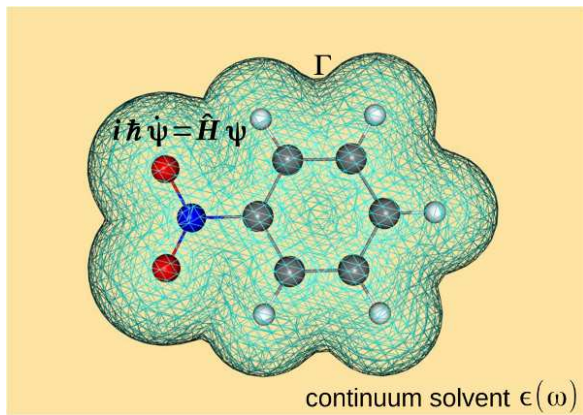}
\caption{\label{pcm_sketch} Schematic representation of the Nitrobenzene molecule embedded in a continuum solvent with a frequency-dependent dielectric constant $\epsilon(\omega)$. $\Gamma$ denotes the surface of the cavity hosting the solute system which is treated quantum-mechanically.}
\end{center}
\end{figure}

A widely used approach to describe molecules in condensed phase consists in the so called focussed models where the system is
divided into parts which are treated at different levels of accuracy. \cite{canuto_solvation_book} An important subset within
these models is constituted by the different formulations of the Polarizable Continuum Model (PCM)
\cite{tomasi_chem_rev_2005} comprising the original dielectric PCM (D-PCM) \cite{ref_pcm_gaussian,miertus_pcm_1982}, the
conductor-like screening model (C-PCM) \cite{dpcm_cossi_jcp_1998, cosmo_klamt_jcs_1993} and the integral equation formalism
(IEF-PCM). \cite{cances_jcp_1997,cances_journal_math_chem_1998} All of them rely on three
basic points as illustrated in Fig. \ref{pcm_sketch}: i) the solute molecule is treated at the quantum mechanical level, ii)
it is hosted by a cavity that exclude the solvent and contains the largest possible amount of the solute charge
density and iii) the solvent is considered a continuous polarizable medium characterized by its frequency-dependent dielectric
function $\epsilon(\omega)$.

PCM implementations are available among the most popular Quantum Chemistry programs using basis sets such as plane waves and
atomic orbitals.\cite{cosmo_klamt_jcs_1993, mennucci_jpcB_1997, scalmani_continuous_pcm_2010, andreussi_jcp_2004} However, in the
last years, with the rapid increase of computational power, real-space finite-difference methods have gained a lot of attention to
perform first-principles electronic structure calculations. \cite{hirose_realspace_book, troullier_pseudo_prl, andrade_jpcm_2012}
They are numerically robust and very accurate since molecular orbitals defined on a spatial grid have the largest flexibility to
take the proper values in both the intra- and the inter-atomic regions. Moreover, the potentials can be directly calculated in
real space without the need of using basis sets, and arbitrary boundary conditions can be used to simulate actual experimental
environments. 

In this paper, we present a methodology to account for solvation effects within real-space and real-time
calculations and demonstrate that solvation energies and solvatochromic shifts can be calculated to chemical accuracy as compared
with state-of-the-art quantum-chemical basis sets methods. Furthermore, we show that this method allows to investigate solvent
effects in the real-time electron dynamics of photoexcited states. In particular, we propose a simple method to regularize the
Coulomb singularity in the solvent reaction potential that may arise for grid points in the simulation domain that are
infinitesimally close to the discretized solute-solvent interface. To that aim, we introduced a set of normalized spherical
Gaussian functions to smooth the discretized polarization charge density. This idea is very similar in spirit to the continuous
surface charge formalism described by Scalmani {\it et al.} in Ref. \onlinecite{scalmani_continuous_pcm_2010} even though the
motivations are different. Within their approach the discretized PCM integral equations have been re-written to avoid
discontinuities in the calculated potential energy surfaces for molecules in solution while in the present work we have focused on
deriving a simple regularized expression for the solvent reaction potential. The
method is implemented in the GPL license {\sc Octopus} code. \cite{octopus_code}

We use the regularized PCM potential to calculate the electrostatic contribution to the solvation
free energies and the optical absorption response for a set of simple organic molecules in water. To calculate the
ground state energy of the solute systems, we use Density Functional Theory (DFT) by solving in real-space the self-consistent
Kohn-Sham (KS) equations. The optical absorption spectra is calculated by exciting the system with a sudden
external perturbation and propagate the KS states in real-time.\cite{yabana_pssB_2006, tddft_book} In this work the real-time
solvent response is calculated in the simplest approximation where the solvent polarization is assumed to equilibrate
instantaneously the propagated electronic density. Different schemes to account for non-equilibrium effects in the real-time
solvent response have been published in the literature \cite{corni_rttddft_2014, liang_rtpcm_2012, nguyen_pcletters_2012,
caricato_jcp_2006, ding_jcp_2015} and can be brought into our formalism. This will be a subject of forthcoming investigation. The
accuracy of our real-space methodology is demonstrated by the excellent agreement of our calculations with the results computed by
using atomic basis sets with the {\sc Gamess} \cite{gamess} software with the advantage of allowing us to explore the excited
states dynamics beyond the linear response regime with no further effort. 

The paper is organized as follows. In Sec. \ref{sec_theory} we explain the theoretical framework used through out the article. We start by introducing the basics of the PCM and the main equation to calculate
the apparent surface charges by using the integral equation formulation of the PCM model. In Sec.
\ref{sec_regularized_pcm} the regularized (singularity-free) solvent reaction potential in real space is derived altogether with the mathematical appendix \ref{regularized_pot_appendix}. The PCM terms entering the
Kohn-Sham Hamiltonian and the expression to calculate the free energy of the solute-solvent system are
obtained in Sec. \ref{subsec_ks_eqs}. Sec. \ref{subsec_rt_pcm} clarifies the approximation used to calculate the solvent
response in real-time. The numerical results calculated by using the new PCM
implementation are discussed and benchmarked in Sec. \ref{sec_application}. Finally, in Sec. \ref{conclusions}, we summarize the main conclusions and comment on possible directions in which this work can be extended.

\section{Theory}
\label{sec_theory}
In the framework of PCM the interaction with the solvent molecules is approximated by the
effective Hamiltonian,
\begin{equation}
 \hat\mathcal{H}= \hat H_\mathrm{M}^0 + \hat V^\mathrm{int},
 \label{effective_hamiltonian}
\end{equation}
where $\hat H_\mathrm{M}^0$ is the electronic Hamiltonian of the molecule in vacuo and $\hat V^\mathrm{int}$ is the solute-solvent
interaction operator expressed in terms of the reaction potentials $V_\mathrm{R}^e({\bf r})$ and $V_\mathrm{R}^n({\bf r})$
accounting for the polarization of the dielectric by the electronic $\rho^e({\bf r})$ and nuclear $\rho^n({\bf r})$ charge
densities of the solute system:

\begin{eqnarray}
\hat V^\mathrm{int} = -\sum_{i=1}^{N_e} [ V_\mathrm{R}^n({\bf r}_i) &+& V_\mathrm{R}^e({\bf r}_i) ]\nonumber\\
&+& \frac{1}{2} \sum_{I=1}^\mathrm{N_{atoms}} Z_I V_\mathrm{R}^n({\bf R}_I),
\label{interaction_term}
\end{eqnarray}
where ${\bf r}_i$, refer to the electrons coordinates and ${\bf R}_I$ and $Z_I$ indicate the positions and the charges of the
atomic nuclei.

The reaction potentials in Eq. (\ref{interaction_term}) can be defined by using an auxiliary apparent surface charge (ASC) $\sigma^{e/n} ({\bf s})$ that spreads on the cavity surface $\Gamma$, \cite{tomasi_chem_rev_2005}
\begin{equation}
V_\mathrm{R}^{e/n}({\bf r}) = \int_\Gamma \frac{\sigma^{e/n}({\bf s})}{\vert {\bf r}- {\bf
s} \vert} d{\bf s}.
\label{v_pcm_asc_def}
\end{equation}
In practice, to solve the PCM equations the continuous surface $\Gamma$ needs to be discretized and approximated by a set of
$T$ finite surface elements commonly referred as {\it tesserae}. As a consequence, the integral in Eq. (\ref{v_pcm_asc_def})
transforms into the discrete sum:
\begin{equation}
V_\mathrm{R}^{e/n}({\bf r}) \approx \sum_{k=1}^T \frac{\sigma^{e/n}({\bf
s}_k) A_k}{\vert {\bf r}- {\bf s}_k \vert} = \sum_{k=1}^T \frac{q^{e/n}({\bf s}_k)}{\vert {\bf
r}- {\bf s}_k \vert},
 \label{v_pcm_discrete}
\end{equation}
where ${\bf s}_k$ and $A_k$ denote, respectively, the representative point and the area of the $k$-th {\it tessera}. In Eq.
(\ref{v_pcm_discrete}) the area of the {\it tesserae} must be small enough to assume a constant value for $\sigma^{e/n}({\bf s})$
within each element which allows to end up with a solvent reaction potential defined by the set of polarization
charges $\{q^{e/n}({\bf s}_k)\}$.

The integral equation formulation of the PCM provides a general framework to calculate these polarization charges as:
\begin{equation}
 q^{e/n}({\bf s}_k) = \sum_{l=1}^T Q({\bf s}_k, {\bf s}_l; \epsilon) V_\mathrm{M}[\rho^{e/n}]({\bf s}_l),
 \label{q_pcm_ief}
\end{equation}
where $Q({\bf s}_k, {\bf s}_l;\epsilon)$ denotes the PCM response matrix which depends on the cavity geometry and the
frequency-dependent dielectric constant of the solvent solution. The expression of the PCM matrix as well as the analytical
formulas to evaluate its matrix elements has been already published elsewhere. \cite{tomasi_chem_rev_2005,stefano_phd_thesis}
As shown in Eq. (\ref{q_pcm_ief}), the polarization charges by the electrons
(nuclei) of the solute molecule are obtained by applying the PCM matrix to the column vector $\{V_\mathrm{M}[\rho^{e/n}]({\bf
s}_l)\}$ containing the electronic (nuclear) contribution to the molecule's electrostatic potential calculated at the
cavity surface as:
\begin{equation}
 V_\mathrm{M}[\rho^{e/n}]({\bf s}_l) = \int d{\bf r} ~ \frac{\rho^{e/n}({\bf r})}{\vert {\bf s}_l - {\bf r} \vert}.
 \label{molecule_elect_pot}
\end{equation}
In Eq. (\ref{molecule_elect_pot}), $\rho^e({\bf r})$ denotes the quantum-mechanical electronic density of the solute system, so
that, $V_\mathrm{M}[\rho^{e}]({\bf s}_l)$ is nothing else than the Hartree potential at the cavity surface and $\rho^n({\bf
r})=\sum_{I=1}^{N_\mathrm{atoms}} Z_I \delta({\bf r} - {\bf R}_I)$ is the classical nuclear density.

\subsection{The solvent reaction potential in real space}
\label{sec_regularized_pcm}
If a real-space grid is used to represent the solvent reaction potential defined by Eq. (\ref{v_pcm_discrete}) a Coulomb
singularity will show up when the coordinate ${\bf r}$ is infinitesimally close to a surface representative point ${\bf
s}_k$. This singularity can lead to significant numerical errors and may represent an important drawback to model solvation
effects in real-space calculations, specially for charged molecules. In particular, for negatively charged systems (anions),
where the solvent response is described by a set of positive polarization charges, the singular nature of the PCM potential in
Eq. (\ref{v_pcm_discrete}) might result in an artificial accumulation of the electrons around the cavity representative points
which is typically manifested in a dramatic overestimation of the solute-solvent interaction energy. The latter effect will be
illustrated and discussed in Section \ref{sec_gs_dft}.

In order to remove such a singularity, we have built a regularized PCM potential by approximating the set of point charges
$\{q({\bf s}_k)\}$  by a set of Gaussian-shaped charge densities $\{\varrho({\bf r},{\bf s}_k)\}$ centered at the
representative points ${\bf s}_k$ whose widths are equal to the areas of the {\it tesserae}:
\begin{equation}
\varrho({\bf r},{\bf s}_k)=\frac{q({\bf s}_k)}{(\pi \alpha A_k)^{3/2}} e^{-\vert {\bf r}-{\bf s}_k \vert^2/(\alpha A_k)}.
 \label{pol_density}
\end{equation}
In Eq. (\ref{pol_density}), $\varrho({\bf r},{\bf s}_k)$ is normalized to the total polarization charge $q({\bf s}_k)$ and
$\alpha$ is a numerical parameter that can be used to modify the width of the charge distribution. In the limit of
$\alpha \rightarrow 0$ the Eq. (\ref{pol_density}) reduces to the case of a point charge located at {\it tessera}
representative point. In the appendix \ref{regularized_pot_appendix} we have calculated the electrostatic potential in
real-space produced by the charge density $\varrho({\bf r},{\bf s}_k)$ and obtained the following expression:
\begin{equation}
v_\mathrm{R}({\bf r},{{\bf s}_k}) = \frac{2q({\bf s}_k)}{\sqrt{\pi \alpha A_k}} \mathcal{G}(\vert {\bf r} - {\bf s}_k
\vert/\sqrt{\alpha A_k}),
\label{pot_realspace_main}
\end{equation}
where $\mathcal{G}(x)$ is a Pad\'e approximant \cite{baker_pade_1996} of order [2/3] defined as:
\begin{equation}
 \mathcal{G}(x)=\frac{1 + p_1 x + p_2 x^2 }{1 + q_1 x + q_2 x^2 + p_2 x^3}.
 \label{pade_eq}
\end{equation}
The coefficients in the latter equation are reported in the appendix \ref{regularized_pot_appendix}. As can be noticed from Eq.
(\ref{pade_eq}) the new potential $v_\mathrm{R}({{\bf s}_k},{\bf r})$ reproduces the $1/\vert {\bf
s}_k - {\bf r} \vert$ dependence for grid points located at large distances from the representative point ${\bf s}_k$ 
and more important, regularizes the Coulomb singularity in real-space at the cavity surface when ${\bf r} =
{\bf s}_k$ which was our primary goal (see Fig. \ref{potential_fit}). Finally, by summing the individual contributions of
all {\it tesserae} we obtain an expression analogous to Eq. (\ref{v_pcm_discrete}) to calculate the regularized solvent reaction
potential as follows:
\begin{equation}
v_\mathrm{R}({\bf r})=\sum_{k=1}^T v_\mathrm{R}({\bf r},{{\bf s}_k}).
\label{v_pcm_regular}
\end{equation}
\subsection{PCM terms in the real-space Kohn-Sham equations}
\label{subsec_ks_eqs}
In the present article solvent effects in the ground-state electronic structure of the solute molecule are accounted for in the framework of Density Functional Theory (DFT) \cite{dft_kohanoff_book}. Therefore, we start from the free energy \cite{comment_free_energy} functional for the solvated system, defined as:
\begin{eqnarray}
\kern - 15 pt G[\rho^e,\rho^n] &=& E^0[\rho^e] \nonumber \\ 
&+& \frac{1}{2}\int d{\bf r} ~ \rho^e({\bf r}) \left\{ v_\mathrm{R}[\rho^e]({\bf r}) + v_\mathrm{R}[\rho^n]({\bf r}) \right\} \nonumber\\
&+& \frac{1}{2}\int d{\bf r} ~\rho^n({\bf r}) \left\{ v_\mathrm{R}[\rho^e]({\bf r}) + v_\mathrm{R}[\rho^n]({\bf r}) \right\},
 \label{free_energy}
\end{eqnarray}
where $E^0[\rho^e]$ is the total energy functional \cite{tddft_ullrich_2011} of the molecule in vacuo and $v_\mathrm{R}({\bf r})$
denotes the solvent reaction potential calculated in Sec. \ref{sec_regularized_pcm} which holds an implicit functional
dependence on the electronic and nuclear charge densities throughout the polarization charges that are calculated
following Eq. (\ref{q_pcm_ief}). If we now take the functional derivative on Eq. (\ref{free_energy}) with respect to the
electronic density $\rho^e({\bf r})$ we obtain for the effective potential of the Kohn-Sham (KS) system the expression:
\begin{eqnarray}
 \kern - 17 ptv_s[\rho^e,\rho^n]({\bf r}) &=& v_s^0[\rho^e]({\bf r}) + \frac{1}{2} \left\{ v_\mathrm{R}[\rho^e]({\bf r}) + v_\mathrm{R}[\rho^n]({\bf r})\right\} \nonumber\\
 &+&\frac{1}{2} \int d{\bf r}^\prime \left\{ \rho^e({\bf r}^\prime) + \rho^n({\bf r}^\prime) \right\} \frac{\delta v_\mathrm{R}[\rho^e]({\bf r}^\prime)}{\delta \rho^e({\bf r}) } 
 \label{v_ks_1},
\end{eqnarray}
with $v_s^0[\rho^e]({\bf r})$ being the corresponding effective potential in vacuo including the exchange-correlation (xc) term.
Formally, the potential $v_\mathrm{R}[\rho^e]({\bf r})$ can be expressed in terms of the electrostatic Green function
\cite{comment_green_function,tomasi_chem_rev_2005}, that is:
\begin{equation}
v_\mathrm{R}[\rho^e]({\bf r}^\prime)=\int d{\bf r}^{\prime\prime} G^\mathrm{R}({\bf r}^\prime,{\bf r}^{\prime\prime})
\rho^e({\bf r}^{\prime\prime}). \label{green_function}
\end{equation}
Substituting Eq. (\ref{green_function}) in the integrand of Eq. (\ref{v_ks_1}), taking the functional derivative
and exploiting the symmetry $G^\mathrm{R}({\bf r}^\prime,{\bf r}^{\prime\prime})=G^\mathrm{R}({\bf r}^{\prime\prime}, {\bf
r}^\prime)$ of the Green function \cite{tomasi_chem_rev_2005}, the expression for the effective potential
$v_\mathrm{s}[\rho^e,\rho^n]({\bf r})$
simplifies as follows:
\begin{equation}
 v_\mathrm{s}[\rho^e,\rho^n]({\bf r}) = v_\mathrm{s}^0[\rho^e]({\bf r}) + v_\mathrm{R}[\rho^e]({\bf r}) + v_\mathrm{R}[\rho^n]({\bf r}).
 \label{v_ks_2}
\end{equation}
The Kohn-Sham molecular orbitals $\varphi_j({\bf r})$ and the eigenvalues $\varepsilon_j$ are obtained by solving the following equation: 
\begin{equation}
\left( -\frac{\nabla^2}{2} + v_\mathrm{s}[\rho^e,\rho^n]({\bf r}) \right) \varphi_j({\bf r}) = \varepsilon_j \varphi_j({\bf r}),
\label{ks_eqs} 
\end{equation}
The KS equations are solved numerically in real-space by using finite differences to approximate the kinetic energy operator and the derivatives of the electronic density that might enter the exchange-correlation potential. Computational details regarding these calculations are reported in Sec. {\ref{sec_comp_details}}.

Eqs. (\ref{ks_eqs}) must be solved iteratively and coupled to the PCM equations (\ref{q_pcm_ief}) and (\ref{v_pcm_regular}) until the equilibrium condition between the solute electronic density and the solvent polarization response has been reached. This is implemented by starting from a guessed electronic density calculated at the level of a linear combination of atomic orbitals (LCAO). The latter density is plugged into Eq. (\ref{q_pcm_ief}) to calculate the initial approximation to the polarization charges $\{q^e({\bf s}_k)\}$. These charges are used to generate the solvent reaction potential $v_\mathrm{R}({\bf r})$ and after solving the KS equations, the new electronic density is obtained from the improved orbitals. This algorithm is repeated so forth until self-consistency. On the other hand, as the nuclei coordinates remain fixed, the solvent response to the nuclear charge density is calculated only once out of the iterative scheme.

Having solved the KS equations, the total free energy for the molecule in solvent can be computed as follows:
\begin{eqnarray}
G[\rho^e,\rho^n] &=& \sum_{j=1}^{N_{occ}} \varepsilon_j -\int d{\bf r}~\rho^e({\bf r}) \left\{v_\mathrm{R}[\rho^e]({\bf r}) + v_\mathrm{R}[\rho^n]({\bf r}) \right\} \nonumber\\
\nonumber\\
&-& \int d{\bf r}~ \rho^e({\bf r}) v_{xc}[\rho^e]({\bf r}) - E_\mathrm{H}[\rho^e] + E_{xc}[\rho^e] \nonumber\\
\nonumber\\
&+& E_\mathrm{int}^\mathrm{el}[\rho^e,\rho^n],
\label{dft_energy}
\end{eqnarray}
where $N_\mathrm{occ}$ is the number of occupied MOs, $E_\mathrm{H}$ is the Hartree energy, $E_{xc}[\rho^e]$ is exchange-correlation energy functional, $v_{xc}[\rho^e]=\delta E_{xc}[\rho^e]/\delta \rho^e({\bf r})$ and $E_\mathrm{int}^\mathrm{el}[\rho^e,\rho^n]$ is the contribution to the free energy due to the electrostatic interaction of the solute molecule with the polarized solvent to be calculated from the following expression:
\begin{eqnarray}
E_\mathrm{int}^\mathrm{el}[\rho^e,\rho^n]=\frac{1}{2} \sum_{k=1}^T &&\left\{V_\mathrm{M}[\rho^e]({\bf s}_k) + V_\mathrm{M}[\rho^n]({\bf s}_k) \right\} \nonumber\\
&&\times \left\{ q^e({\bf s}_k) + q^n({\bf s}_k) \right\}.
 \label{elect_int_energy}
\end{eqnarray}

\subsection{Solvent response in real-time}
\label{subsec_rt_pcm}
To model the electron dynamics of the solute molecule under the influence of an arbitrary time-dependent
external potential, we use the time-dependent Kohn-Sham (TDKS) equations in the adiabatic approximation: \cite{gross_tddft}
\begin{equation}
i\hbar\frac{\partial}{\partial t} \varphi_j({\bf r},t) = \left( -\frac{\nabla^2}{2} + v_\mathrm{s}[\rho^e,\rho^n]({\bf r},t)
\right) \varphi_j({\bf r},t),
\label{td_ks_eqs} 
\end{equation}
where $v_\mathrm{s}[\rho^e,\rho^n]$ is the effective potential defined by Eq. (\ref{v_ks_2}) evaluated in the instantaneous density $\rho^e({\bf r},t)$ which is obtained by propagating in real-time the ground state KS orbitals $\varphi({\bf r},t=0)$. 

Under these conditions, the polarization charges induced by the electrons of the solute molecule become also time-dependent and
should be computed, as it is done for ground state calculations, coupled to the quantum-mechanical TDKS equations. However, in the
static case the solvent polarization caused by the solute molecule is in equilibrium with the ground state density but, in
general, as the time-dependent electronic density varies faster as compare with the typical timescales of the solvent relaxation
processes, the actual polarization will lag behind the changing density. 

In this work, we have performed real-time TDDFT (RT-TDDFT) calculations for simple organic molecules in
solvent to benchmark our real-space PCM implementation. In particular, we compare the optical response of the solute molecules
obtained from the dynamic polarizability tensor \cite{tddft_book} with linear-response TDDFT (LR-TDDFT)
calculations performed in the frequency domain by using Gaussian basis sets. These calculations were carried out by using equal
values for the dynamic and static dielectric constants. Under this assumption the calculation of the time-dependent apparent
charges simplifies to the expression: \cite{corni_rttddft_2014}
\begin{equation}
 q^e({\bf s}_k,t)=\sum_{l=1}^T Q({\bf s}_k,{\bf s}_l;\epsilon_0) \mathrm{V}_\mathrm{M}[\rho^e]({\bf s}_l,t).
 \label{td_pcm_eqs}
\end{equation}
Eq. (\ref{td_pcm_eqs}) is the simplest expression to calculate the solvent response in real-time where the dielectric
polarization is assumed to equilibrate instantaneously the evolved electronic density. This approximation, despite it
neglects non-equilibrium effects due to the delayed components of the solvent polarization, constitutes a very good approach
suitable to treat molecules in weakly polar solvents characterized in most cases by similar values of the static and dynamic
dielectric constants.

\section{Implementation and Results}
\label{sec_application}
\begin{figure}[b]
\begin{center}
\includegraphics[width=1 \linewidth,angle=0]{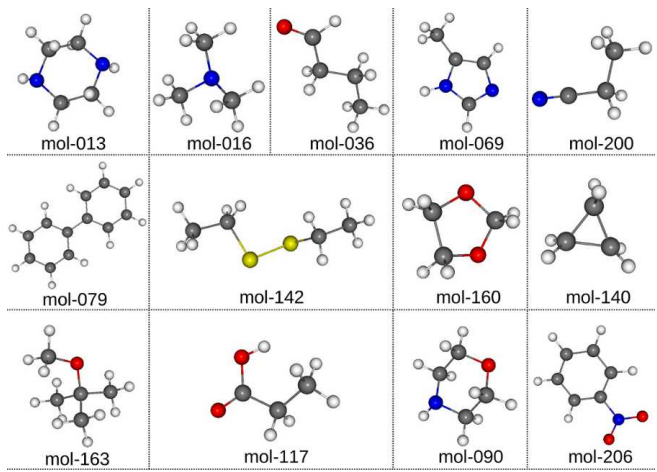}
\caption{\label{mol_structures} Optimized structures of the molecules used to benchmark the PCM implementation. The color code
for the elements is red for O, blue for N, grey for C, yellow for S and white for H. The chemical names of these molecules are
reported in the supplemental figures S1-S4. \cite{supplemental_material}}
\end{center}
\end{figure}

We have implemented the PCM equations recalled above as a new module within the {\sc Octopus} code. \cite{octopus_code}
In this section, we report on numerical calculations carried out to test and benchmark the PCM implementation described in Sec.
\ref{sec_regularized_pcm}. We have performed calculations of the ground state energies and the optical absorption
response for a set of organic molecules (see Fig. \ref{mol_structures}) that were recently used for benchmarking
purposes in Ref. \onlinecite{andreussi_2012_revised} in water solution. The static and time-dependent KS equations
(\ref{ks_eqs}) and (\ref{td_ks_eqs}), have been solved in real-space and real-time by using {\sc Octopus}.
Numerical details about these calculations are explained in the following section. The calculated
solvation free energies and solvatochromic shifts for selected molecules are discussed in sections \ref{sec_gs_dft} and
\ref{sec_tddft_absorption}, respectively, and compared with similar calculations performed with the
{\sc Gamess} software. \cite{gamess} All the results reported in the present paper were computed by using the PBE
generalized gradient approximation to the exchange and correlation energy functionals. \cite{pbe_functional_original,
pbe_functional_erratum}

\subsection{Computational details}
\label{sec_comp_details}
The molecular structures shown in Fig. \ref{mol_structures} were optimized in gas phase by DFT calculations, as
implemented in {\sc Gamess} by using the double zeta basis set 6-31G(d). \cite{basis_631Gd_pople, emsl_gaussian_bsets} See
supplemental material at [URL will be inserted by AIP] for the geometries used in the reported calculations.
\cite{supplemental_material}

The calculated geometries were inputed to {\sc Octopus} to perform DFT and TDDFT calculations in vacuo and solvent. The
simulation box containing the real-space domain was constructed by adding spheres around each atomic center of radius $5~
\mathrm{\AA}$. Calculations performed with larger radii showed that this value is sufficient to get total and molecular
orbitals energies converged for all the studied molecules. Ground state calculations were carried out by using two different
values for the uniform spacing, $0.15~\mathrm{\AA}$ and $0.19~\mathrm{\AA}$, between each grid point of the generated mesh.
Atomic pseudo-potentials have been used to model the core electrons for all the species. They were generated consistently 
with the {\sc Ape} \cite{ape_code} code by using PBE approximation to the xc functional.

To calculate the absorption spectra of some of the molecules shown in Fig. \ref{mol_structures}, we propagated in real-time the KS
system from its ground state after applying, at $t=0$, a weak delta pulse of a dipole electric field. Real time propagations were
performed along the three polarization directions of the applied field by using a time step of $\Delta t=1.7$ attosecond (as)
for a total simulation time of 20 femtoseconds (fs). The enforced time-reversal symmetry method, \cite{propagators_castro_2004,
propagators_tddft_octopus} as implemented in {\sc Octopus}, was used to propagate the time-dependent KS orbitals. The
optical absorption strength was calculated by using the imaginary part of the diagonal component of the dynamic polarizability
tensor defined in the frequency domain by Fourier transforming the time-dependent electric dipole. \cite{tddft_book} The time
integral over the total simulation time has been exponentially filtered with a damping factor of $0.13$ eV.

The cavity surface $\Gamma$ hosting the solute molecule is built as a union of interlocking spheres centered on all atoms but
hydrogens. \cite{pascual_gepol_1994} We have used for the radii of these spheres the values $2.4
~\mathrm{\AA}$ for C, $1.8~\mathrm{\AA}$ for O, $1.9~\mathrm{\AA}$ for N and $2.02~\mathrm{\AA}$ for S. \cite{frediani_jcp_2004}
The elements of the PCM response matrix $Q({\bf s}_k,{\bf s}_l;\epsilon)$ are calculated by using Eqs. (A9, A15) and the first
terms of Eqs. (A10, A14), as derived in the appendix of Ref. \onlinecite{adelgado_jcp_2013}. We used
$\epsilon=78.39$ for the static dielectric constant of water. \cite{dielectric_constants} Polarization charges
are calculated with Eq. (\ref{q_pcm_ief}) in terms of the PCM response matrix and the molecule's electrostatic potential at the
cavity surface. The nuclear contribution to the latter potential can be directly evaluated by using Eq.
(\ref{molecule_elect_pot}). On the other hand, to calculate the electronic contribution at the cavity surface, we
perform a trilinear interpolation by using the values of the Hartree potential at the grid points defining the vertices of the
cube containing the {\it tessera} representative point (see inset in Fig. \ref{fig_singularity_test}). 

Once the total polarization charges are computed, the solvent reaction potential in real space, entering the Kohn-Sham equations,
is calculated through Eq. (\ref{v_pcm_regular}). All the results presented along the paper have been obtained by using
$\alpha=1$ in Eq. (\ref{pot_realspace_main}) where the PCM potential is regularized in terms of the areas of the {\it tesserae}.
In the supplemental figures S1-S4 \cite{supplemental_material} we report the weak dependence of the solute-solvent interaction
energy for different values of the parameter $\alpha$ entering the regularized solvent reaction potential.

We have used {\sc Gamess} to benchmark the solvation free energies and the optical absorption spectra obtained with {\sc Octopus}
for the investigated molecules in water. The same cavity surface and PCM response matrix were used within both codes. Total
energies and excited state calculations with {\sc Gamess} were performed by using the triple zeta basis set 6-311+(d,p).
\cite{emsl_gaussian_bsets}

\subsection{Electrostatic contribution to the solvation free energy}
\label{sec_gs_dft}
At the level of ground state calculations, the solvation free energy is an important magnitude to characterize quantitatively
the solute-solvent interaction in thermodynamic equilibrium. In general, PCM allows also for the introduction of non-electrostatic
terms \cite{tomasi_chem_rev_2005, andreussi_2012_revised} in the solute Hamiltonian given by Eq. (\ref{effective_hamiltonian})
which are not considered in the present discussion. In what follows we focus our analysis on the calculated results for the
electrostatic contribution to the solvation free energy,
\begin{equation}
 \Delta G_\mathrm{el} = G[\rho_\mathrm{solv}^e,\rho^n]-E[\rho_\mathrm{vac}^e,\rho^n],
 \label{solv_ener_def}
\end{equation}
where $G[\rho_\mathrm{solv}^e,\rho^n]$ is given by Eq. (\ref{free_energy}) and $E[\rho_\mathrm{vac}^e,\rho^n]$ is the total
energy of the isolated solute in vacuum. 

\begin{figure}[t]
\begin{center}
\includegraphics[width=1 \linewidth,angle=0]{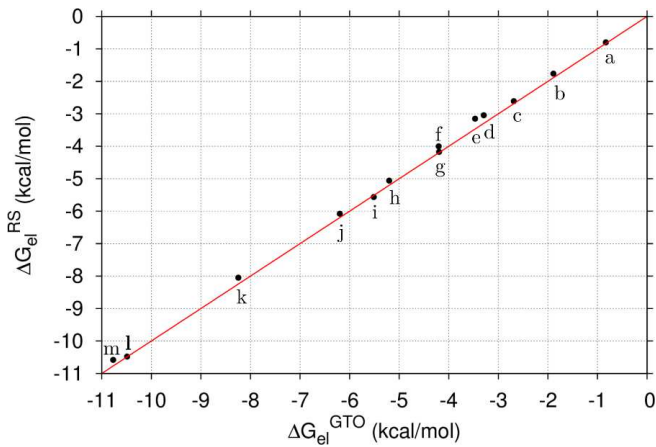}
\caption{\label{fig_solv_ener} Comparison of the electrostatic contribution to the solvation free energies calculated in
real-space (RS) with {\sc Ooctopus} with similar results obtained with {\sc Gamess} by using Gaussian type orbitals (GTO) for the
molecules shown in Fig. \ref{mol_structures}. The superimposed letters are used to identify each molecule in Table
\ref{table_free_energies}.}
\end{center}
\end{figure}

In Fig. \ref{fig_solv_ener}, real-space calculations for $\Delta G_\mathrm{el}$ are compared with the corresponding results
obtained by using a Gaussian basis set for the whole set of molecules shown in Fig. \ref{mol_structures} in water solution.
These results were obtained by using a grid spacing of $0.19~\mathrm{\AA}$ and $\alpha=1$ to generate the
regularized solvent reaction potential (Eq.(\ref{v_pcm_regular})) in real space. From the figure, an excellent agreement between
the different calculations is evident. As a matter of fact, all plotted circles hardly deviate from the diagonal,
showing maximum differences of about $0.3~\mathrm{kcal/mol}$. The calculated free energies are also reported in
Table \ref{table_free_energies} for the sake of completeness. 
\begin{table}[t]
\caption{Electrostatic contribution to the solvation free energies in kcal/mol plotted in Fig. \ref{fig_solv_ener}. Real-space
(RS) calculations were performed by using a grid spacing of $0.19 \mathrm{\AA}$ with {\sc Octopus}. The analogous results by
using Gaussian type orbitals (GTO) were calculated with {\sc Gamess}.}
\label{table_free_energies}
\begin{tabular}{cccc}
\hline\noalign{\smallskip} 
 & molecule & \hspace{1.4 cm} $\Delta G_\mathrm{el}^\mathrm{RS}$ & \hspace{1.4 cm} $\Delta
G_\mathrm{el}^\mathrm{GTO}$ \\
\noalign{\smallskip}\hline\noalign{\smallskip}
a) & mol-140 & \hspace{1.1 cm} $-0.79$  & \hspace{1.1 cm} $-0.83$  \\
b) & mol-016 & \hspace{1.1 cm} $-1.76$  & \hspace{1.1 cm} $-1.88$  \\
c) & mol-163 & \hspace{1.1 cm} $-2.60$  & \hspace{1.1 cm} $-2.68$  \\
d) & mol-142 & \hspace{1.1 cm} $-3.04$  & \hspace{1.1 cm} $-3.29$  \\
e) & mol-079 & \hspace{1.1 cm} $-3.14$  & \hspace{1.1 cm} $-3.46$  \\
f) & mol-036 & \hspace{1.1 cm} $-4.17$  & \hspace{1.1 cm} $-4.19$  \\
g) & mol-160 & \hspace{1.1 cm} $-4.00$  & \hspace{1.1 cm} $-4.20$  \\
h) & mol-206 & \hspace{1.1 cm} $-5.05$  & \hspace{1.1 cm} $-5.20$  \\
i) & mol-200 & \hspace{1.1 cm} $-5.56$  & \hspace{1.1 cm} $-5.51$  \\
j) & mol-090 & \hspace{1.1 cm} $-6.07$  & \hspace{1.1 cm} $-6.19$  \\
k) & mol-013 & \hspace{1.1 cm} $-8.05$  & \hspace{1.1 cm} $-8.24$  \\
l) & mol-117 & \hspace{1.1 cm} $-10.48$ & \hspace{1.1 cm} $-10.49$ \\
m) & mol-069 & \hspace{1.1 cm} $-10.58$ & \hspace{1.1 cm} $-10.77$ \\
\noalign{\smallskip}\hline\hline\noalign{\smallskip}
\end{tabular}
\end{table}
\begin{figure}[b!]
\begin{center}
\includegraphics[width=1 \linewidth,angle=0]{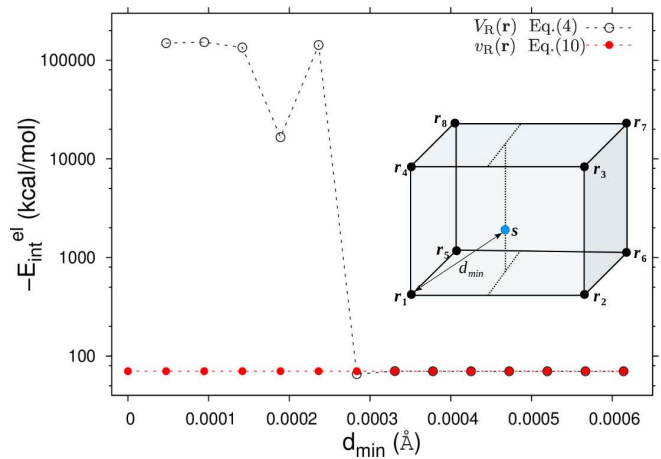}
\caption{\label{fig_singularity_test} Calculated electrostatic interaction energy for a chlorine anion in water for different
values of the distance between the {\it tessera} representative point ${\bf s}$ and the real-space point ${\bf r}_1$. $0.15
\mathrm{\AA}$ was used as grid spacing. The $\Gamma$ surface in this case is defined by a sphere of radius $2.17 \mathrm{\AA}$.}
\end{center}
\end{figure}
We have performed the same calculations without regularizing the PCM potential by using Eq. (\ref{v_pcm_discrete}) to evaluate the
solvent reaction field in real-space and obtained very similar results as it is shown in Fig. S5 of the supplemental
material.\cite{supplemental_material}
This suggests that numerical problems due to the Coulomb singularity are unlikely and also that the value
$\alpha=1$ is a very good choice for the regularization. The same analysis has been carried out by using
a smaller grid spacing of $0.15~\mathrm{\AA}$ and we found, again, the same trend and accuracy. 

However, there may be a threshold for the minimum distance between a {\it tessera} and a grid point for which numerical
instabilities should appear. In order to illustrate this possibility, we consider one of the worst case scenario.   
In Fig. \ref{fig_singularity_test} we plot the interaction energy given by Eq. (\ref{elect_int_energy}) calculated for the
chlorine anion in water as a function of the distance $d_\mathrm{min}$ between the representative point ${\bf s}$ at the cavity
surface and a grid point ${\bf r}_1$. This is sketched by the inset superimposed to Fig. \ref{fig_singularity_test} where $\{{\bf
r}_1,..,{\bf r}_8\}$ label the vertices of the cube surrounding the {\it tessera}. Results are presented for two cases: open
circles correspond to calculations performed with a PCM potential given by Eq. (\ref{v_pcm_discrete}); solid circles plot the
interaction energies computed by using the regularized potential as defined in Eq. (\ref{v_pcm_regular}). As it appears from the
plot, there is a critical distance (in this case $\mathrm{d}_\mathrm{min} \approx 2.8 \times 10^{-4}~ \mathrm{\AA}$) below 
which the singular nature of the Coulomb potential $V_\mathrm{R}({\bf r})$ manifests dramatically. However, as it can be noticed
from the same figure, such numerical problems are completely absent in the calculations carried out with the regularized PCM
potential predicting accurate energies for all distances, even when the {\it tessera} coordinates fully overlaps with the grid
point ($\mathrm{d}_\mathrm{min}=0$). It is noteworthy that this result was also obtained by setting $\alpha = 1$.

\subsection{Solvent effects in the optical absorption response}
\label{sec_tddft_absorption}
In this section we report and discuss the optical response of selected molecules: piperazine
(mol-013), trimethylaminne (mol-016), biphenyl (mol-079) and nitrobenzene (mol-206).
\begin{figure}[b!]
\begin{center}
\includegraphics[width=0.97 \linewidth,angle=0]{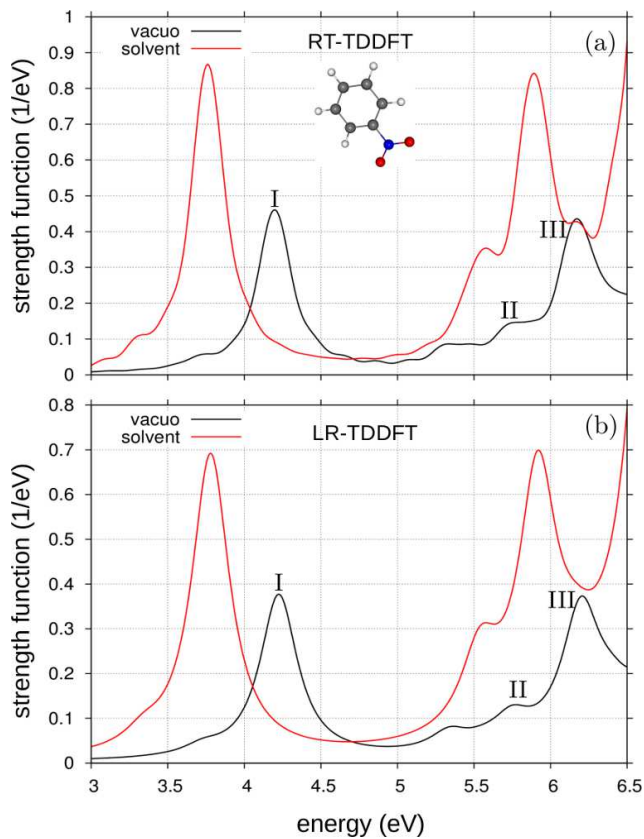}
\caption{\label{fig_spectra_mol_206} Absorption spectra of nitrobenzene molecule (mol-206) calculated with (a) RT-TDDFT by using
{\sc Octopus} and (b) LR-TDDFT with {\sc Gamess}.}
\end{center}
\end{figure}
The absorption spectra of these systems have been calculated in gas phase and in water by solving in real-time and real-space the
TDKS (Eq. (\ref{td_ks_eqs})) and PCM (Eq. (\ref{td_pcm_eqs})) equations simultaneously, as implemented in our code of choice, {\sc
Octopus}. The computed spectra have been benchmarked with LR-TDDFT calculations by using Gaussian type orbitals as implemented in
{\sc Gamess}.

In Fig. \ref{fig_spectra_mol_206} we plot the calculated strength function versus the excitation energy for the nitrobenzene
molecule. The spectra obtained by doing RT-TDDFT calculations are shown in the upper panel
while results obtained within linear response approximation are reported in the lower panel. Notice from this figure that the
calculated spectra, both in vacuo and in solvent, with RT-TDDFT  and LR-TDDFT are almost identical, that is, the main absorption
bands appears at the same excitation energies with similar relative intensities. In particular, we focus
our analysis on how the solvatochromic shifts of the main absorption bands obtained by real-time propagation compare with the
analogous result obtained by using a Gaussian basis set; this comparison offers a direct measure of the
quality of the PCM implementation presented along this paper. To that aim, we have labeled with roman numerals the absorption
peaks in the gas phase spectrum that can be clearly identified in the solvated one for the calculated molecules.
See supplemental material at [URL will be inserted by AIP] where the absorption spectra of the molecules piperazine
(mol-013), trimethylaminne (mol-016) and byphenil (mol-079) are plotted in Figs. S6-S8.
\cite{supplemental_material}
\begin{figure}[b!]
\begin{center}
\includegraphics[width=1 \linewidth,angle=0]{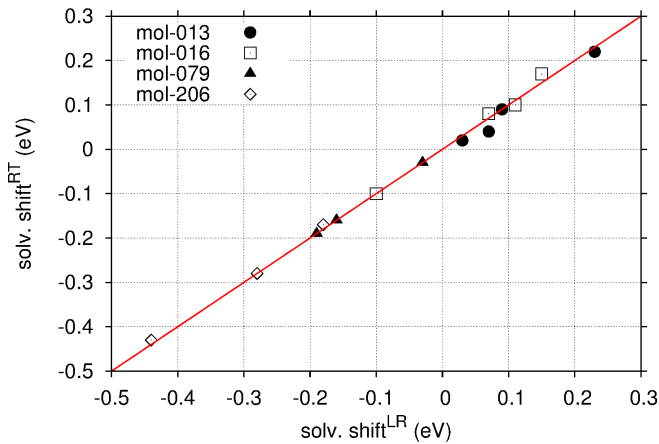}
\caption{\label{fig_solv_shifts} Solvatochromic (solv.) shifts of the main absorption bands of the molecules
mol-013, mol-016, mol-079 and mol-206 in water. Real-time (RT) TDDFT calculations are correlated with
analogous results obtained in the linear response (LR) approximation. See also Table \ref{table_solv_shifts}.}
\end{center}
\end{figure}

\begin{table}[t]
\caption{Solvatochromic shifts in eV of the main absorption bands for the selected molecules in water obtained from real-time (RT)
and linear response (LR) TDDFT calculations.}
\label{table_solv_shifts}
\begin{tabular}{cccc}
\hline\noalign{\smallskip} 
molecule & \hspace{0.3 cm} peak & \hspace{0.3 cm} $\mathrm{solv.~shift}^\mathrm{RT}$ & \hspace{0.3 cm}
$\mathrm{solv.~shift}^\mathrm{LR}$ \\
\noalign{\smallskip}\hline\noalign{\smallskip}
mol-013 & I    & $0.22$ & $0.23$ \\
        & II   & $0.02$ & $0.03$ \\
        & III  & $0.09$ & $0.09$ \\
        & IV   & $0.04$ & $0.07$ \\
\noalign{\smallskip}\hline\noalign{\smallskip}
mol-016 & I    & $0.10$  & $0.11$  \\
        & II   & \hspace{-0.35 cm} $-0.10$ & \hspace{-0.35 cm} $-0.10$ \\
        & III  & $0.17$  & $0.15$  \\
        & IV   & $0.08$  & $0.07$  \\
\noalign{\smallskip}\hline\noalign{\smallskip}
mol-079 & I    & \hspace{-0.35 cm} $-0.19$ & \hspace{-0.35 cm} $-0.19$ \\
        & II   & \hspace{-0.35 cm} $-0.03$ & \hspace{-0.35 cm} $-0.03$ \\
        & III  & \hspace{-0.35 cm} $-0.16$ & \hspace{-0.35 cm} $-0.16$ \\
\noalign{\smallskip}\hline\noalign{\smallskip}
mol-206 & I    & \hspace{-0.35 cm} $-0.43$ & \hspace{-0.35 cm} $-0.44$ \\
        & II   & \hspace{-0.35 cm} $-0.17$ & \hspace{-0.35 cm} $-0.18$ \\
        & III  & \hspace{-0.35 cm} $-0.28$ & \hspace{-0.35 cm} $-0.28$ \\
\noalign{\smallskip}\hline\hline\noalign{\smallskip}
\end{tabular}
\end{table}

In Fig. \ref{fig_solv_shifts} we correlate the solvatochromic shifts for the main absorption bands of the investigated
molecules as obtained from RT-TDDFT calculations with the corresponding values calculated with {\sc
Gamess} in the linear response approach. The plotted values report for the energy differences taken at the absorption maxima
between the solvated and gas phase spectra of the calculated molecules. As it turns out from this plot, RT-TDDFT calculations
predict shifted excitation energies due to the dynamical solvent polarization in perfect agreement with linear
response calculations performed by using Gaussian atomic functions. We have found a maximum deviation of $0.03$ eV from the
diagonal which corresponds to the solvatochromic shift of the highest-energy band of piperazine. For the sake of completeness we
report in Table \ref{table_solv_shifts} the values of the solvatochromic shifts plotted in Fig. \ref{fig_solv_shifts}.

\section{Conclusions}
\label{conclusions}
 
In this work we presented a new methodology to account for solvation effects in the electronic and optical
properties of molecular systems by using density functional approximations in real-space and real-time. We have used
the Integral Equation Formalism of the Polarizable Continuum Model to calculate the apparent charges at the cavity
surface hosting the solute molecule. To prevent numerical instabilities in real-space calculations due to Coulomb
singularities in the solvent reaction potential we introduced a set of Gaussian functions, centered at the {\it tesserae}
representative points, to distribute the polarization charges within the area of each surface element.
This allowed us to derive a simple analytical expression to evaluate the PCM potential over the whole simulation domain. We used
the regularized solvent potential to do ground state calculations by solving the self-consistent Kohn-Sham (KS) equations by using
the PBE generalized gradient approximation to the xc energy. The new PCM implementation was applied to compute in
real-space the solvation free energies of a set of organic molecules in water by using the {\sc Octopus} code. For benchmarking
proposes we compared with similar calculations by using Gaussian atomic orbitals performed with {\sc Gamess} and found an
excellent agreement between both results that showed maximum differences of about 0.3 kcal/mol.

We have also extended the PCM equations to the real-time domain to investigate solvent effects in the electron dynamics
under the influence of a time-dependent external field. We calculated the optical spectra of
selected molecules in water by propagating in real-time the KS orbitals and the PCM apparent
charges, after perturbing the molecule with a weak instantaneous dipole. The obtained solvatochromic shifts for the
main absorption bands compare extremely well with the analogous results calculated in the frequency domain within the
linear-response approximation by using {\sc Gamess}. At present, the real-time solvation effects has been included by
assuming that solvent polarization equilibrates instantaneously the evolved electronic density. This approach, despite it neglects
non-equilibrium effects in the dielectric response, is a very good approximation to investigate molecules in weakly polar
solvents.

The methodological developments and the results presented in this paper set the grounds for further extensions to model solvation
effects in the framework of real-space and real-time electronic structure calculations by using density functional approximations.
For instance, the inclusion of non-equilibrium effects to describe the solvent response in real-time can be incorporated into our
computational machinery as proposed in Ref. \onlinecite{corni_rttddft_2014}. This would contribute
to evaluate dielectric effects more rigorously in the time-resolved optical spectroscopy of molecular systems in solution
to probe ultrafast charge transfer and non-linear optical phenomena. Moreover, this work
constitutes a good starting point to add solvation effects to model the coupled electron-nuclei dynamics in
the Ehrenfest approximation as implemented in the {\sc Octopus} code.
\cite{ehrenfest_andrade_2009,carlo_nature_2013}

\begin{acknowledgments}
We acknowledge financial support from the EU FP7 project ``CRONOS'' (grant agreement:280879). S.P. acknowledges financial support from the European Community's FP7 Marie Curie IIF MODENADYNA Grant Agreement:623413. The authors thanks the scientific staff of CINECA for support. All the calculations were carried out in the framework of the ISCRA-C project OCT-PCM.
\end{acknowledgments}

\appendix

\section{Regularization of the solvent reaction potential in real-space}
\label{regularized_pot_appendix}
In real-space calculations, it is evident from Eq. (\ref{v_pcm_discrete}) that the solvent reaction potential $V_\mathrm{R}({\bf
r})$ shows a singularity for grid points that are infinitesimally close to the solute cavity surface, that is, for ${\bf r}
\rightarrow {\bf s}_k$. To regularize the PCM potential at this limit, we distributed isotropically the polarization
charges within each {\it tessera} by using a Gaussian function centered at the representative point ${\bf
s}_i$ and width equals to the area $A_k$,  
\begin{equation}
 \varrho({\bf r},{\bf s}_k)=N e^{-\vert {\bf r}-{\bf s}_k \vert^2/(\alpha A_k)}.
 \label{gaussian_density}
\end{equation}
The parameter $\alpha$ in Eq. (\ref{gaussian_density}) has been introduced to modify the width of the Gaussian smearing out the
polarization charges. The constant $N$ can be obtained from the normalization condition:
\begin{equation}
 \int d{\bf r}~\varrho({\bf r},{\bf s}_k) = N 4 \pi \int_0^\infty dx ~ x^2 e^{-\beta_k x^2}=q({\bf s}_k),
 \label{norma_condition}
\end{equation}
where $\beta_k=1/(\alpha A_k)$ and $x=\vert {\bf r} - {\bf s}_k \vert$. Making use of the parametric integral,
\begin{equation}
 \int_0^\infty dx ~ e^{-\beta_k x^2} = \frac{1}{2} \sqrt{\frac{\pi}{\beta_k}},
 \label{parametric_integral}
\end{equation}
we obtain the normalized charge density:
\begin{equation}
\varrho({\bf r},{\bf s}_k)=\frac{q({\bf s}_k)}{(\pi \alpha A_k)^{3/2}} e^{-\vert {\bf r}-{\bf s}_k \vert^2/(\alpha A_k)}.
 \label{normalized_density}
\end{equation}

On the other hand, by using the Gauss law and the spherical symmetry of $\varrho$, the radial component of the electric field at
a certain distance $x_k=\vert {\bf r} - {\bf s}_k \vert$ from the representative point, can be calculated as follows:

\begin{equation}
E(x_k) = \frac{1}{4\pi\epsilon_0} \frac{\tilde q(x_k)}{x_k^2},
 \label{electric_field}
\end{equation}
where $\tilde q(x_k)$ is the amount of polarization charge inside the spherical surface of radius $x_k$ to be calculated by
integrating the charge density in Eq. (\ref{normalized_density}),
\begin{equation}
\tilde q(x_k) = \frac{4\pi q({\bf s}_k)}{(\pi \alpha A_k)^{3/2}} \int_0^{x_k} dy ~ y^2 e^{-y^2/(\alpha A_k)}.
 \label{partial_charge}
\end{equation}
By changing to the variable $t=y^2/(\alpha A_k)$ the latter integral rewrittes as follows:
\begin{eqnarray}
 \tilde q(x_k)&=&\frac{2q({\bf s}_k)}{\sqrt{\pi}} \left[ \int_0^\infty dt~ t^{1/2} e^{-t} - \int_{x_k^2/(\alpha A_k)}^\infty dt~
t^{1/2} e^{-t} \right]
\nonumber \\
 &=&\frac{2q({\bf s}_k)}{\sqrt{\pi}} \left[ \Gamma(3/2) - \Gamma(3/2,x_k^2/(\alpha A_k)) \right],\label{partial_charge_1}
\end{eqnarray}
where $\Gamma(s)$ and $\Gamma(s,x)$ denote the complete and upper incomplete gamma functions \cite{table_integrals_jeffrey_2007},
respectively. Now, with an analytic expression for $\tilde q(x_k)$ we can calculate the electrostatic potential generated by
$\varrho({\bf r},{\bf s}_k)$ at any point in real space by integrating Eq. (\ref{electric_field}),
\begin{eqnarray}
\kern-25pt v&(&{\bf r},{\bf s}_k)=-\frac{2q({\bf s}_k)}{\sqrt{\pi}}  \nonumber\\
&\cdot& \int_\infty^{\vert {\bf r} - {\bf s}_k \vert } dy
~\left\{ \frac{\Gamma(3/2)-\Gamma(3/2,y^2/(\alpha A_k))}{y^2}\right\}. 
 \label{pot_realspace}
\end{eqnarray}
The integrand of the latter equation can be further simplified by changing to the dimensionless variable $x^2=y^2/(\alpha A_k)$
and scaling out the factor $\alpha A_k$ to the integration limit, that is: 
\begin{equation}
v({\bf r},{\bf s}_k) = \frac{2q({\bf s}_k)}{\sqrt{\pi \alpha A_k}} \mathcal{F}(\vert {\bf r} - {\bf s}_k
\vert/\sqrt{\alpha A_k}),
\label{pot_realspace_1}
\end{equation}
with the function $\mathcal{F}(x)$ defined as,
\begin{equation}
\mathcal{F}(x)= \int_x^\infty dx^\prime ~ \left\{
\frac{\Gamma(3/2)-\Gamma(3/2,{x^\prime}^2)}{{x^\prime}^2} \right\}.
\label{pot_realspace_2}
\end{equation}
\begin{figure}[h]
\begin{center}
\includegraphics[width=1.0\linewidth,angle=0]{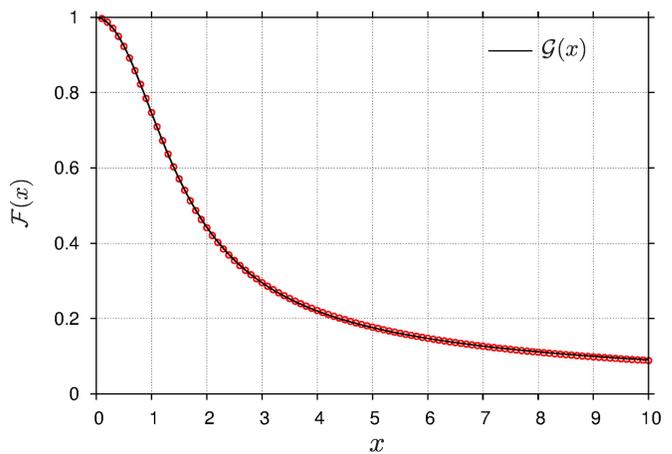}
\caption{\label{potential_fit} Calculated values for the function $\mathcal{F}(x)$ (solid circles) obtained from Eq.
(\ref{pot_realspace_2}). The Pad\'e approximant $\mathcal{G}(x)$, used to fit the asymptotic behavior and functional
dependence of $\mathcal{F}(x)$, is plotted as a solid line.}
\end{center}
\end{figure}
The argument of function $\mathcal{F}(x)$ takes values between $0$, for the case ${\bf r}={\bf s}_k$ and $\infty$. In Fig.
\ref{potential_fit} we plot the calculated values for the integral in Eq. (\ref{pot_realspace_2}) as function of $x$ in the
interval between 0 and 10 atomic units. Notice that the potential resulting from Eq. (\ref{pot_realspace_1})
reproduce the expected $1/r$ dependence at large distances from the {\it tesserae} representative points, while in the limit of
${\bf r}={\bf s}_k$ regularize the Coulomb singularity to the constant value $2q({\bf s}_k)/\sqrt{\pi \alpha A_k}$ which was our
primary goal. These asymptotic limits as well as the functional dependence of the $\mathcal{F}(x)$ can be perfectly
fitted by using the Pad\'e approximant: \cite{baker_pade_1996}
\begin{equation}
 \mathcal{G}(x)=\frac{1 + p_1 x + p_2 x^2 }{1 + q_1 x + q_2 x^2 + p_2 x^3},
 \label{pade_fit}
\end{equation}
with the coefficients $p_1=0.119763$, $p_2=0.205117$, $q_1=0.137546$ and $q_2=0.434344$. The excellent agreement between
the proposed fit and the exact results calculated from the integral in Eq. (\ref{pot_realspace_2}) is shown in Fig.
\ref{potential_fit}.


\begin{thebibliography}{51}
\bibitem [{com({\natexlab{b}})}]{comment_green_function}%
  \BibitemOpen
  \href@noop {} \bibinfo {note} {$G^\mathrm{R}({\bf
  r},{\bf r}^\prime) = G^\mathrm{tot}({\bf r},{\bf r}^\prime) -
  G^\mathrm{0}({\bf r},{\bf r}^\prime)$, where $G^\mathrm{tot}({\bf r},{\bf
  r}^\prime)$ solves the equation $-\epsilon\nabla^2_{\bf r}
  G^\mathrm{tot}({\bf r},{\bf r}^\prime) = 4 \pi \delta({\bf r} - {\bf
  r}^\prime)$ with the proper boundary conditions on the cavity surface and
  $G^\mathrm{0}({\bf r},{\bf r}^\prime) = 1 / \vert {\bf r} - {\bf r}^\prime
  \vert$}\BibitemShut {NoStop}%
\end{thebibliography}
\end{document}